# Modeling of causality with metamaterials

Igor I. Smolyaninov

*Department of Electrical and Computer Engineering, University of Maryland, College Park, MD 20742, USA*

**Hyperbolic metamaterials may be used to model a 2+1 dimensional Minkowski spacetime in which the role of time is played by one of the spatial coordinates. When a metamaterial is built and illuminated with a coherent extraordinary laser beam, the stationary pattern of light propagation inside the metamaterial may be treated as a collection of particle world lines, which represents a complete "history" of this 2+1 dimensional spacetime. While this model may be used to build interesting spacetime analogs, such as metamaterial "black holes" and "big bang", it lacks causality: since light inside the metamaterial may propagate back and forth along the "timelike" spatial coordinate, events in the "future" may affect events in the "past". Here we demonstrate that a more sophisticated metamaterial model may fix this deficiency via breaking the mirror and temporal (PT) symmetries of the original model and producing one-way propagation along the "timelike" spatial coordinate. Resulting 2+1 Minkowski spacetime appears to be causal. This scenario may be considered as a metamaterial model of the Wheeler-Feynman absorber theory of causality.**

Recent advances in electromagnetic metamaterials and transformation optics gave rise to such fascinating devices as perfect lenses [1], invisibility cloaks [2-7], perfect absorbers [8], and numerous other novel electromagnetic devices. On the other hand, development of these techniques has led to considerable progress in modelling unusual spacetime geometries, such as black holes [9-12], wormholes [13,14], Alcubierre warp drive [15], and spinning cosmic strings [16]. Hyperbolic metamaterials are especially interesting in this respect since extraordinary rays in a hyperbolic metamaterial behave as particle world lines in a three dimensional (2+1) Minkowski spacetime [17,18]. When this spacetime is "curved", metamaterial analogues of black holes [19] and the big bang [18] may be created. Unfortunately, such metamaterial-based Minkowski spacetime analogues described so far are limited in one very important respect. They lack causality. The role of a timelike coordinate in these models is played by one of the spatial coordinates. For example, $z$-coordinate may play the role of time. Since light inside the metamaterial may propagate back and forth along this timelike $z$ coordinate, events in the "future" may affect events in the "past". Here we demonstrate that a more sophisticated metamaterial model may fix this deficiency via breaking the mirror and temporal (PT) symmetries of the original model and producing one-way propagation along the "timelike" $z$ coordinate. Resulting effective 2+1 dimensional Minkowski spacetime appears to be "causal": all events inside such a metamaterial may be arranged into causal sets.

Let us begin with a brief summary of refs.[17-19], which demonstrated that extraordinary rays in a hyperbolic metamaterial behave as particle world lines in a 2+1 dimensional Minkowski spacetime, and that one of spatial coordinates may become "timelike" in a hyperbolic metamaterial. Let us assume that a non-magnetic uniaxial metamaterial may be described by dielectric permittivities $\varepsilon_x = \varepsilon_y = \varepsilon_1 > 0$ and $\varepsilon_z = \varepsilon_2 < 0$, and that this behaviour holds in some frequency range around $\omega = \omega_0$. We will consider propagation of extraordinary light in such a metamaterial (vector $\vec{E}$ of the extraordinary light is parallel to the plane defined by the $k$–vector of the wave and the optical axis of the metamaterial). We will define a "scalar" extraordinary wave function as $\varphi=E_z$ so that the ordinary portion of the electromagnetic field does not contribute to $\varphi$. Since hyperbolic metamaterials exhibit considerable temporal dispersion, we will work in the frequency domain and write the macroscopic Maxwell equations as [20]

$$\frac{\omega^2}{c^2}\vec{D}_\omega = \vec{\nabla}\times\vec{\nabla}\times\vec{E}_\omega \text{ and } \vec{D}_\omega = \vec{\varepsilon}_\omega\vec{E}_\omega \quad, \tag{1}$$

which results in the following wave equation for $\varphi_\omega$ :

$$-\frac{\omega^2}{c^2}\varphi_\omega = \frac{\partial^2\varphi_\omega}{\varepsilon_1\partial z^2} + \frac{1}{\varepsilon_2}\left(\frac{\partial^2\varphi_\omega}{\partial x^2} + \frac{\partial^2\varphi_\omega}{\partial y^2}\right) \tag{2}$$

When the hyperbolic metamaterial is illuminated by coherent CW laser field at frequency $\omega_0$, spatial distribution of the extraordinary field at this frequency is defined by the 3D Klein-Gordon equation for a massive scalar field $\varphi_\omega$ :

$$-\frac{\partial^2\varphi_\omega}{\varepsilon_1\partial z^2} + \frac{1}{|\varepsilon_2|}\left(\frac{\partial^2\varphi_\omega}{\partial x^2} + \frac{\partial^2\varphi_\omega}{\partial y^2}\right) = \frac{\omega_0^2}{c^2}\varphi_\omega = \frac{m^{*2}c^2}{\hbar^2}\varphi_\omega \tag{3}$$

in which spatial coordinate $z=\tau$ behaves as a "timelike" variable. Thus, eq.(3) describes world lines of massive particles which propagate in a flat 2+1 dimensional Minkowski spacetime [17-19]. Moreover, if a dipole source oscillating at frequency $\omega_0$ is placed inside the hyperbolic metamaterial, its radiation pattern looks like a light cone in Minkowski spacetime (see Fig.1). The latter result may be understood qualitatively from the following simple argument. At large wave vectors the right hand side of eq.(3) may be neglected. Thus, any differentiable function $F(\xi)$ gives rise to a solution of eq.(3) of the form

$$\varphi_\omega = aF(\vec{\rho}+\sqrt{-\varepsilon_1/\varepsilon_2}z) + bF(\vec{\rho}-\sqrt{-\varepsilon_1/\varepsilon_2}z) \,, \tag{4}$$

where $\vec{\rho}$ is the radius vector in the *(x,y)* plane, and *a* and *b* are arbitrary coefficients. As can be seen from eq.(4) and Fig.1, the dipole radiates equally into the "past" and the "future" light cones. Thus, our model 2+1 dimensional Minkowski spacetime lacks causality. This lack of causality is a natural consequence of the PT symmetry of our hyperbolic metamaterial: it does not change under the z→-z and *t*→-*t* transformations. This feature of our model differs drastically from physics of real 4D Minkowski spacetime, where dipoles are assumed to radiate only in the positive *t* direction. (Note however that this notion is not universally accepted – see for example the Wheeler-Feynman absorber theory described in refs.[21,22]).

In order to design a causal model of 2+1 dimensional Minkowski spacetime we need to break PT symmetry in such a way that only the "future" light cone will remain intact, while radiation into the "past" light cone will be eliminated. As a result, signals will propagate only in the positive *z* direction. Examples of such one-way propagation due to broken PT symmetry may be found in numerous optical and microwave systems [23-25]. Typically, they involve utilization of magneto-electric materials and ferrites [26], nonlinear optical effects or explicit time-dependent behavior of the system [27]. Let us demonstrate how a model of causal 2+1 dimensional Minkowski spacetime may be designed based on similar approaches.

High frequency macroscopic electrodynamics of metamaterials may be described using two equivalent languages [20]. We can either introduce magneto-electric moduli relating (D,H) and (E,B) pairs:

$$\vec{D} = \vec{\varepsilon}\vec{E} + \vec{\alpha}\vec{B}, \qquad (5)$$

$$\vec{H} = \vec{\mu}^{-1}\vec{B} + \vec{\beta}\vec{E},$$

or assume that $\vec{D} = \vec{\varepsilon}\vec{E}$ and $\vec{H} = \vec{B}$, while tensor $\vec{\varepsilon}(\omega_0, \vec{k})$ exhibits linear (odd) terms in spatial dispersion:

$$\varepsilon_{ij}(\omega_0, \vec{k}) = \varepsilon_{ij}(\omega_0, 0) + \gamma_{ijl}^{(1)} k_l + \gamma_{ijlm}^{(2)} k_l k_m + ..., \qquad (6)$$

Connection between these two descriptions is easy to establish for time harmonic plane waves, since

$$\vec{B} = \frac{c}{i\omega} rot\vec{E} = \frac{c}{\omega} \left[ \vec{k} \times \vec{E} \right], \qquad (7)$$

If nonlinear optical effects need to be taken into account, description in terms of spatial dispersion becomes more convenient since we can treat both spatial dispersion and nonlinear effects in a unified fashion. We can write

$$D_i = \chi_{ij}^{(1)} E_j + \chi_{ijl}^{(2)} E_j E_l + \chi_{ijlm}^{(3)} E_j E_l E_m + ..., \qquad (8)$$

where the nonlinear susceptibility terms $\chi$ may exhibit spatial dispersion. Nonlinear optical interactions may contribute to effective $\vec{\varepsilon}(\omega_0, \vec{k})$ via multiple terms in (8). For example, if we assume that

$$\vec{E} = \vec{E}(\omega_0) + \vec{E}(\omega), \qquad (9)$$

it is clear that the third order nonlinear effects always contribute to effective $\vec{\varepsilon}(\omega_0, \vec{k})$. Second order nonlinear effects may also strongly contribute to $\vec{\varepsilon}(\omega_0, \vec{k})$ if phase matching conditions for either sum or difference frequency generation are satisfied inside the metamaterial. For example, if $\omega=2\omega_0$ and phase-matching exists between the first and second harmonic along some particular direction inside the metamaterial, there will be strong $k$-dependent coupling between $\omega_0$ and $2\omega_0$ fields.

Let us demonstrate that presence of linear (odd) $k$ terms in effective spatial dispersion $\vec{\varepsilon}(\omega_0, \vec{k})$ may lead to one-way "causal" propagation of signals along the "timelike" $z$ coordinate in a hyperbolic metamaterial. The dispersion law of extraordinary photons in the absence of spatial dispersion

$$\frac{k_z^2}{\varepsilon_1} + \frac{k_x^2 + k_y^2}{\varepsilon_2} = \frac{\omega^2}{c^2}, \qquad (10)$$

(where $\varepsilon_x = \varepsilon_y = \varepsilon_1 > 0$ and $\varepsilon_z = \varepsilon_2 < 0$) is shown in Fig.2(a). Note that two hyperbolic branches of the dispersion law, which are symmetric with respect to $z \to -z$ transformation, are separated by large $2\varepsilon_1^{1/2}\omega/c$ gap in $k$ space. Therefore, we may infer several suitable functional forms of spatial dispersion $\varepsilon_{ij}(\omega_0, \vec{k})$, which eliminate the lower hyperbolic branch of the dispersion law at $k_z < 0$ as shown in Fig.2(b). Since off-diagonal terms of $\varepsilon_{ij}(\omega_0, \vec{k})$ are typically much smaller than the diagonal terms, for the sake of simplicity let us consider a situation where off-diagonal terms may be neglected. As will be demonstrated below, this situation may be realized via nonlinear optical effects. One option presented in Fig.2(c) describes a hyperbolic "$\varepsilon$ near zero" (ENZ) material in which $\varepsilon_1 > 0$ does not exhibit much spatial dispersion, while $\varepsilon_2$ has near zero value at $k=0$. ENZ materials necessarily exhibit strong spatial dispersion [28], and linear terms in $\varepsilon_{ij}(\omega_0, \vec{k})$ must dominate at small wave vectors $k_z$. Since on the other hand $\varepsilon_{ij}(\omega_0, \vec{k})$ must be bound in the $k_z \to \pm\infty$ limits, functional shape of $\varepsilon_{ij}(\omega_0, \vec{k})$ presented in Fig.2(c) must be typical in ENZ hyperbolic metamaterials exhibiting linear spatial dispersion. As a result, we have an unusual situation in which $\varepsilon_2$ is negative at $k_z > \varepsilon_1^{1/2}\omega/c$ and positive at $k_z < -\varepsilon_1^{1/2}\omega/c$. This functional behaviour of $\varepsilon_2$ eliminates lower hyperbolic branch of dispersion relation as shown in Fig.2(b). Radiation pattern of a dipole source placed inside such a medium has been calculated using COMSOL Multiphysics 4.2 solver. As expected, only the "future" light cone remains in the radiation pattern (see Fig.3(a)).

Another possible scenario realizing "causal" hyperbolic metamaterial may be based on nonreciprocal directional dichroism [24,29]. This scenario is presented in Fig.2(d). Breaking mirror z→-z symmetry may lead to asymmetric behavior of *Im(ε)*, and hence anisotropic propagation losses. Due to large gap $2\varepsilon_1^{1/2}\omega/c$ separating positive and negative hyperbolic branches of the dispersion law of extraordinary photons, nonreciprocal directional dichroism [29] may lead to suppression of one of these branches resulting in "causal" behavior of the dipole radiation pattern inside the metamaterial. Such radiation pattern calculated using COMSOL Multiphysics 4.2 solver is presented in Fig.3(b). Note that this scenario may be considered as a metamaterial model of the Wheeler-Feynman absorber theory of causality [21,22]: the dipole radiates into both "future" and "past" light cones. However, radiation into the "past" is strongly absorbed. We should also note that real hyperbolic metamaterials always have large losses. Therefore, in the presence of linear spatial dispersion separation of scenarios considered in Fig.2 (c) and (d) is somewhat artificial. In reality, both scenarios happen at the same time.

Let us consider potential experimental realizations of these scenarios using metamaterials. A simplest experimental arrangement realizing a "causal" hyperbolic metamaterial is shown in Fig.4. Let us assume that a circular polarized ordinary wave at frequency $2\omega_0$ propagates inside this metamaterial in the positive $z$ direction. Thus, our geometry explicitly breaks PT symmetry, while metamaterial geometry remains uniaxial, and we may neglect off-diagonal terms in the $\vec{\varepsilon}(\omega_0,\vec{k})$ tensor. The metamaterial consists of a metal wire array with volume fraction $\alpha$ and dielectric constant $\varepsilon_m$ inside a nonlinear dielectric medium having dielectric constant $\varepsilon_d$. Diagonal components of the dielectric tensor of such a metamaterial for small $\alpha$ have been calculated in [30]:

$$\varepsilon_2 = \alpha\varepsilon_m + (1-\alpha)\varepsilon_d \text{, and } \varepsilon_1 \approx \varepsilon_d\frac{(1+\alpha)}{(1-\alpha)} \qquad (11)$$

The ENZ conditions are realized if $\alpha \approx -\varepsilon_d / \varepsilon_m$. Under these conditions we may disregard spatial dispersion of $\varepsilon_1$. On the other hand, spatial dispersion of $\varepsilon_2$ will be very pronounced and dominated by terms linear in $k$. If the frequency dispersion of $\varepsilon_d$ may be disregarded, the $2\omega_0$ ordinary wave and all the extraordinary waves at frequency $\omega_0$ propagating in the positive z direction ($k_z>0$) having small $k_x$ and $k_y$ are phase-matched (see eq.(10) and Fig.4). Such an arrangement attenuates propagation losses of $k_z>0$ extraordinary photons at frequency $\omega_0$, while not affecting $k_z<0$ photons, leading to strong nonreciprocal directional dichroism of $\varepsilon_2$ similar to one shown in Fig.2(d). On the other hand, according to eq.(8) real part of effective $\varepsilon_2$ will also acquire a $k$-dependent nonlinear correction

$$\varepsilon_2^{NL} \sim \chi_{zzx}^{(2)} E_x^{2\omega} + \chi_{zzy}^{(2)} E_y^{2\omega}, \qquad (12)$$

which peaks around the phase matching condition $k_\omega^e = 1/2 k_{2\omega}^o$, where indices "e" and "o" indicate extraordinary and ordinary wave, respectively. Thus, the causal hyperbolic metamaterial scenario presented in Fig.2(c) may also be realized by suitable choice of sign of $\chi^{(2)}$ of the nonlinear dielectric material.

In conclusion, we have presented a metamaterial model of "causal" 2+1 dimensional Minkowski spacetime in which the role of time is played by one of the spatial coordinates. Breaking the mirror and temporal (PT) symmetries of the hyperbolic metamaterial produces one-way propagation of the extraordinary photons along the "timelike" spatial coordinate. As a result, metamaterial becomes "causal", and all events inside such a metamaterial may be arranged into causal sets. Our design may be considered as a metamaterial model of the Wheeler-Feynman [21,22] absorber theory of causality. Obvious limitation of our model comes from the fact that in order to preserve "3D Lorentz symmetry" of the original model we need to avoid off-diagonal terms in the dielectric tensor. Thus, we cannot employ such more common solutions as PT symmetry breaking using ferrites or magneto-electric materials in our design [26]. As a result, we use a nonlinear material, whereas causality of wave propagation in the real 4D Minkowski space-time does not rely on non-linear mechanisms.

**References**

[1] J. B. Pendry, "Negative Refraction Makes a Perfect Lens", *Phys. Rev. Letters* **85**, 3966-3969 (2000).

[2] J. B. Pendry, D. Schurig, D.R. Smith, "Controlling electromagnetic fields", *Science* **312**, 1780-1782 (2006).

[3] U. Leonhardt, "Optical conformal mapping", *Science* **312**, 1777-1780 (2006).

[4] D. Schurig, J.J. Mock, B.J. Justice, S.A. Cummer, J.B. Pendry, A.F. Starr, D.R. Smith, "Metamaterial electromagnetic cloak at microwave frequencies", *Science* **314**, 977-980 (2006).

[5] I.I. Smolyaninov, Y.J. Hung, C.C. Davis, "Two-dimensional metamaterial structure exhibiting reduced visibility at 500 nm", *Optics Letters* **33**, 1342-1344 (2008)

[6] I.I. Smolyaninov, V.N. Smolyaninova, A.V. Kildishev, V.M. Shalaev, "Anisotropic metamaterials emulated by tapered waveguides: application to electromagnetic cloaking", *Phys. Rev. Letters* **102**, 213901 (2009).

[7] V. N. Smolyaninova, I. I. Smolyaninov, H. K. Ermer, "Experimental demonstration of a broadband array of invisibility cloaks in the visible frequency range", New *Journal of Physics* **14**, 053029 (2012).

[8] N.I. Landy, S. Sajuyigbe, J.J. Mock, D.R. Smith, W.J. Padilla, W. J., "Perfect metamaterial absorber", *Phys. Rev. Lett*. **100**, 207402 (2008).

[9] I.I. Smolyaninov, "Surface plasmon toy-model of a rotating black hole", *New Journal of Physics* **5**, 147 (2003).

[10] I.I. Smolyaninov, "Critical opalescence in hyperbolic metamaterials", Journal of Optics 13, 125101 (2011).

[11] D.A. Genov, S. Zhang, X. Zhang, "Mimicking celestial mechanics in metamaterials", *Nature Physics*, **5**, 687-692 (2009).

[12] E.E. Narimanov, A.V. Kildishev, "Optical black hole: Broadband omnidirectional light absorber", *Appl. Phys. Lett.* **95**, 041106 (2009).

[13] A. Greenleaf, Y. Kurylev, M. Lassas, G. Uhlmann, "Electromagnetic wormholes and virtual magnetic monopoles from metamaterials", *Phys. Rev. Lett.* **99**, 183901 (2007).

[14] I.I. Smolyaninov, "Metamaterial "Multiverse"", *Journal of Optics* **13,** 024004 (2010).

[15] I.I. Smolyaninov, "Metamaterial-based model of the Alcubierre warp drive", *Phys. Rev. B* **84**, 113103 (2011).

[16] T. G. Mackay, A. Lakhtakia, "Towards a metamaterial simulation of a spinning cosmic string", *Phys. Lett. A* **374**, 2305-2308 (2010).

[17] I.I. Smolyaninov, E.E. Narimanov, "Metric signature transitions in optical metamaterials", *Phys. Rev. Letters* **105**, 067402 (2010).

[18] I.I. Smolyaninov and Y.J. Hung, "Modeling of time with metamaterials", *JOSA B* **28**, 1591-1595 (2011).

[19] I.I. Smolyaninov, E. Hwang, and E.E. Narimanov, "Hyperbolic metamaterial interfaces: Hawking radiation from Rindler horizons and spacetime signature transtions", *Phys. Rev. B* **85**, 235122 (2012).

[20] L.Landau, E.Lifshitz, Electrodynamics of Continuous Media (Elsevier, 2004).

[21] J.A. Wheeler, R.P. Feynman, "Interaction with the absorber as the mechanism of radiation", *Rev. Mod. Phys.* **17**, 157–161 (1945).

[22] J.A. Wheeler, R.P. Feynman, "Classical electrodynamics in terms of direct interparticle action". *Rev. Mod. Phys.* **21**, 425–433 (1949).

[23] B.L. Johnson, H.-H. Shiau, "Guided magneto-plasmon polaritons in thin films: non-reciprocal propagation and forbidden modes", *J.Phys.: Condens. Matter* **20**, 335217 (2008).

[24] I. Kezsmarki, N. Kida, H. Murakawa, S. Bordacs, Y. Onose, Y. Tokura, "Enhanced directional dichroism of terahertz light in resonance with magnetic excitations of the multiferroic $Ba_2CoGe_2O_7$ oxide compound", *Phys. Rev. Letters* **106**, 057403 (2011).

[25] V.N. Gridnev, "Low frequency behavior of optical spatial dispersion effects", *Physics of the Solid State* **43**, 682-684 (2001).

[26] B. Lax and K.J. Button, Microwave Ferrites and Ferrimagnetics (McGraw-Hill, New York, 1962).

[27] G. Shvets, "Not every exit is an entrance", *Physics* **5**, 78 (2012).

[28] M.G. Silveirinha, "Nonlocal homogenization model for a periodic array of $\epsilon$-negative rods", *Phys. Rev. E* **73**, 046612 (2006).

[29] S. Miyahara, N. Furukawa, "Nonreciprocal directional dichroism and toroidalmagnons in helical magnets", *J. Phys. Soc. Jpn*. **81**, 023712 (2012).

[30] R. Wangberg, J. Elser, E.E. Narimanov, V.A. Podolskiy, "Nonmagnetic nanocomposites for optical and infrared negative-refractive-index media", *JOSA B* **23**, 498-505 (2006).

**Figure Captions**

**Figure 1.** Radiation pattern of a dipole source placed inside a hyperbolic metamaterial calculated using COMSOL Multiphysics 4.2 solver. Radiation pattern looks like a light cone in a 2+1 dimensional Minkowski spacetime in which spatial z coordinate plays the role of a "timelike" coordinate.

**Figure 2**. (a) Dispersion law of a hyperbolic metamaterial, which does not exhibit spatial dispersion. (b) Desired dispersion law in a "causal" hyperbolic metamaterial. (c) Functional behavior of $\varepsilon_2$ as a function of $k_z$ in an $\varepsilon$ near zero (ENZ) hyperbolic metamaterial having nonzero linear spatial dispersion. This behavior results in "causal" propagation of signals inside the metamaterial. (d) Functional behavior of Im($\varepsilon_2$) as a function of $k_z$ due to nonreciprocal directional dichroism. Such behavior also results in "causal" signal propagation.

**Figure 3**. Radiation pattern of a dipole source placed inside "causal" hyperbolic metamaterials calculated using COMSOL Multiphysics 4.2 solver. Only the "future" light cone remains in the radiation pattern. Case (a) corresponds to scenario presented in Fig.2(c), while case (b) corresponds to scenario presented in Fig.2(d).

**Figure 4.** Possible experimental realization of "causal" hyperbolic metamaterial: ENZ hyperbolic metamaterial operated at frequency $\omega_0$ is made of metal wire array inside a nonlinear dielectric. The metamaterial is illuminated along $z$-direction with a circular polarized ordinary wave at frequency $2\omega_0$.

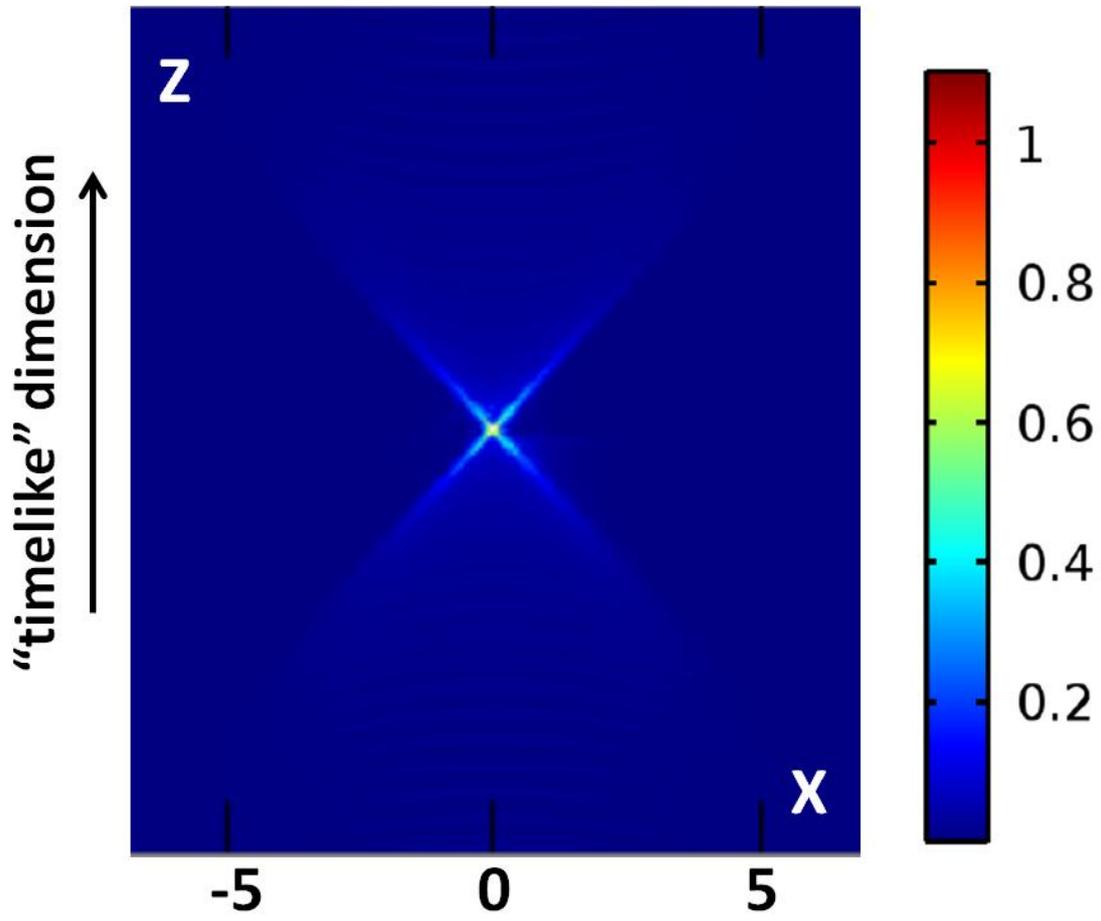

Fig.1

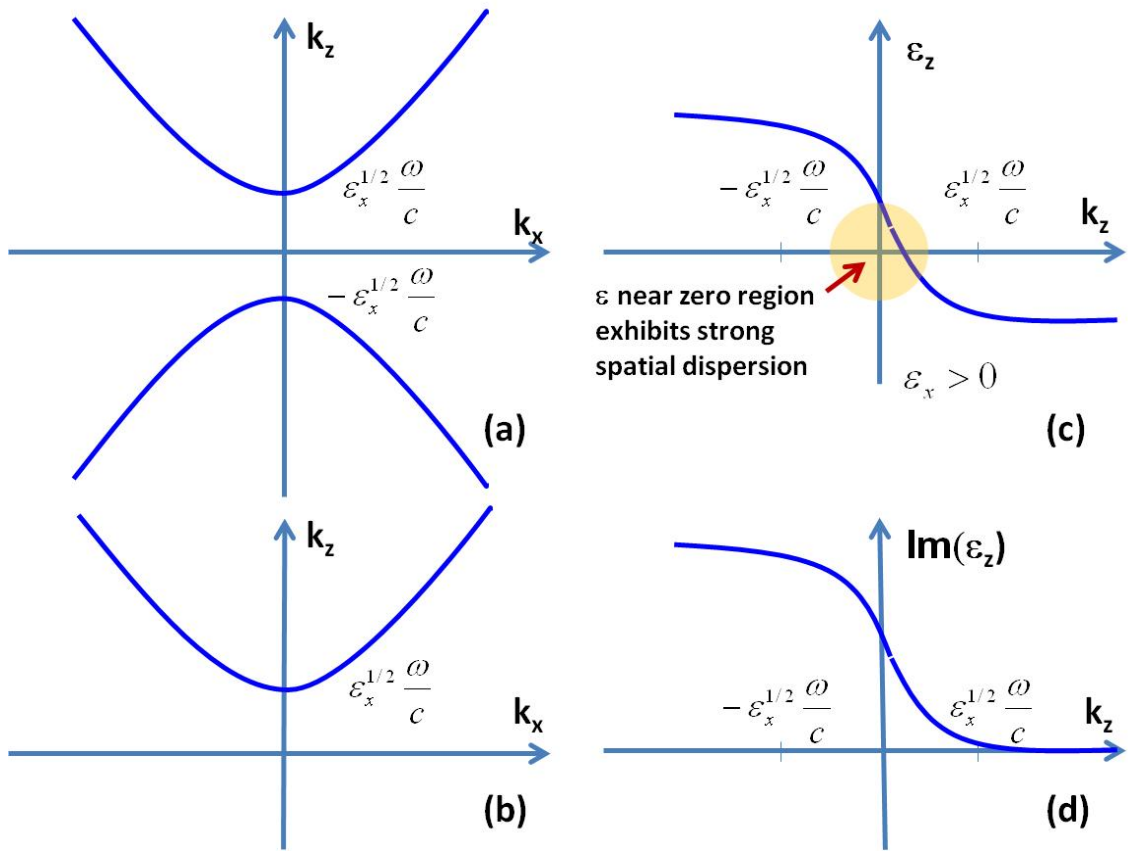

Fig. 2

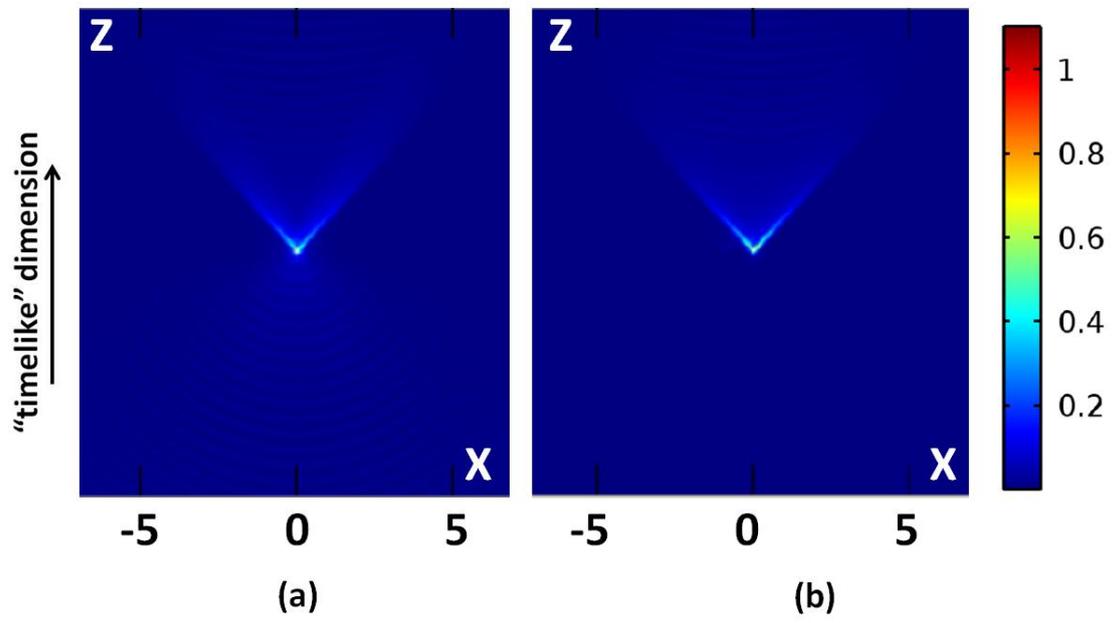

Fig. 3

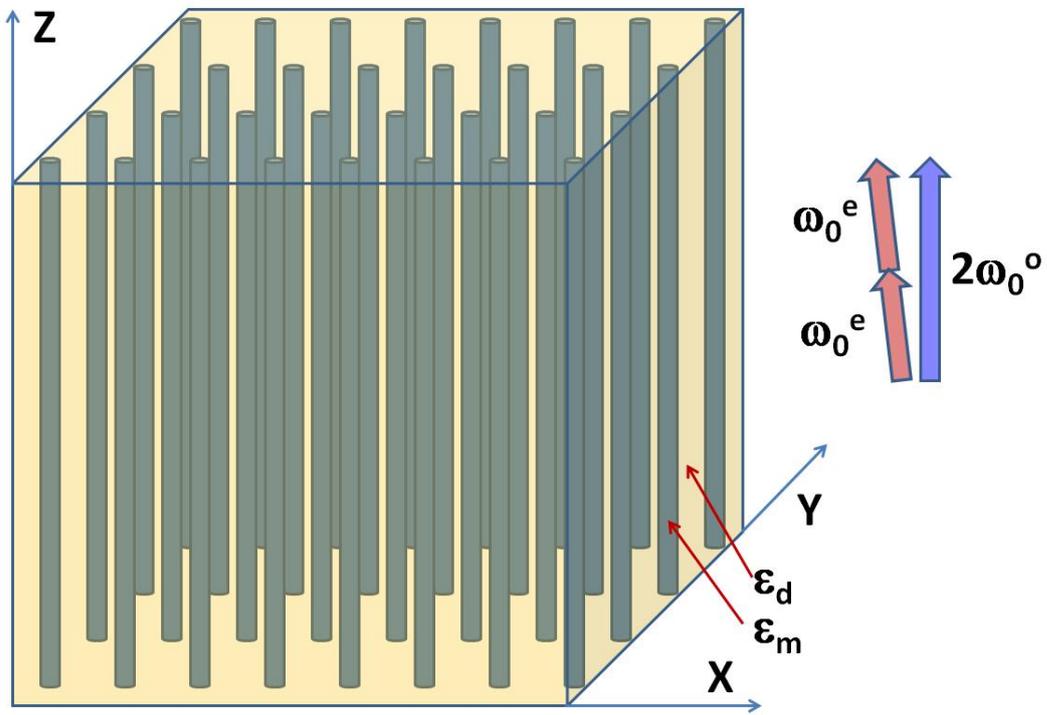

Fig. 4